\documentclass[aps,prl,twocolumn,showpacs,
10pt,superscriptaddress,preprintnumbers,nofootinbib]{revtex4-1}
\usepackage{amsmath,bm,braket}
\usepackage{slashed}
\usepackage{color}
\usepackage{graphics}
\usepackage{epsfig}

\allowdisplaybreaks

\def\L{\mathcal{L}}
\def\Ma{M}
\def\siga{\frac{d\sigma}{d\Ma}}
\def\sigat{\frac{d\tilde{\sigma}}{d\Ma}}

\def\Nat{\tilde{N}}
\def\fig#1{fig.~{\ref{#1}}}

\def\eqn#1{eq.~(\ref{#1})}

\begin{document}

\preprint{SLAC-PUB-15463}

\title{Bounding the Higgs Boson Width Through Interferometry}

\author{Lance J. Dixon}
\affiliation{SLAC National Accelerator Laboratory, Stanford University, 
Stanford, CA 94309, USA}
\author{Ye Li}
\affiliation{SLAC National Accelerator Laboratory, Stanford University, 
Stanford, CA 94309, USA}

\begin{abstract}
We study the change in the di-photon invariant mass distribution for
Higgs boson decays to two photons, due to interference between the Higgs 
resonance in gluon fusion and the continuum background amplitude 
for $gg\to\gamma\gamma$. 
Previously, the apparent Higgs mass was found to shift by around 100 MeV
in the Standard Model in the leading order approximation, which may potentially
be experimentally observable.  We compute the next-to-leading order QCD 
corrections to the apparent mass shift, which reduce it by about 40\%.
The apparent mass shift may provide a way to measure, or at
least bound, the Higgs boson width at the Large Hadron Collider
through ``interferometry.''
We investigate how the shift depends on the Higgs width, in a
model that maintains constant Higgs boson signal yields.
At Higgs widths above 30 MeV the mass shift is over 200 MeV and 
increases with the square root of the width.  The apparent mass
shift could be measured by comparing with the $ZZ^*$ channel,
where the shift is much smaller.  It might be possible to
measure the shift more accurately by exploiting its strong dependence
on the Higgs transverse momentum.
\end{abstract}


\maketitle


\section{Introduction}

Recently, experiments at the Large Hadron Collider (LHC)
have discovered a new boson with a mass
around 125 GeV~\cite{CMSHiggs,AtlasHiggs}, whose properties are roughly
consistent with those predicted for the Standard Model (SM)
Higgs boson.   It is now crucial to determine its properties as accurately
as possible.  The Higgs boson is dominantly produced by gluon
fusion through a top quark loop.  Its decay to two photons,
$H\to\gamma\gamma$, provides a very clean signature for probing Higgs
properties, including its mass.  However, there is also a large
continuum background to its detection in this channel.
It is important to study how much the coherent interference between
the Higgs signal and the background could affect distributions in
diphoton observables, and possibly use it to constrain Higgs properties.

The interference of the Higgs boson with $gg\to \gamma\gamma$ was first 
studied for an intermediate Higgs mass boson~\cite{Dicus1987fk}.
For the experimentally relevant case of a light, narrow-width Higgs
boson, it is fruitful to divide the interference contribution into two parts,
proportional to the real and imaginary parts of the Higgs boson's
Breit-Wigner propagator, respectively.  The diphoton invariant-mass
distribution for the real-part interference is odd around the Higgs
mass. It contributes negligibly to the experimentally observed cross
section, which is integrated over the narrow lineshape.
The imaginary part interferes constructively or destructively with
the signal distribution at the Higgs mass.  For a light SM Higgs boson, 
the imaginary part vanishes at leading order in the zero quark mass
limit~\cite{Dicus1987fk}.
The dominant contribution comes from the two-loop $gg\to\gamma\gamma$
amplitude and gives only a few percent suppression of the
rate~\cite{Dixon2003yb}.  However, the real part interference is
affected by finite detector resolution, which smears the
diphoton invariant mass distribution and causes a sizable shift
in the apparent Higgs mass peak, as pointed out in
ref.~\cite{Martin2012xc} and further studied in
refs.~\cite{deFlorian2013psa,Martin2013ula}.

In this letter, we calculate the dominant
next-to-leading order (NLO) QCD
corrections to the interference, and study the dependence of the mass shift
on the acceptance cuts.  We further argue that the interference effect, 
especially the mass shift, can be used to bound experimentally,
or possibly even measure, the Higgs width fairly directly, for widths well
below the experimental mass resolution at the LHC.
Such a measurement would complement even more
direct measurements of the Higgs width at future colliders
such as the ILC~\cite{Richard2007ru,Peskin2012we} or a muon
collider~\cite{Han2012rb,Conway2013lca}, but might be accomplished
much earlier.

Indirect bounds on the Higgs width at the LHC have also been given, based
on global analyses of various Higgs decay
channels~\cite{Dobrescu2012td,Djouadi2013qya,CMSPAS}. However, in these
analyses it is
impossible to decouple the Higgs width from the couplings without
a further assumption, because the Higgs signal strength is
always given by the product of squared couplings for Higgs production
and for decay, divided by the Higgs total width $\Gamma_H$.
Typically the further assumption is that the Higgs coupling
to electroweak vector bosons does not exceed the SM value.
For example, a recent CMS analysis making this assumption obtained a
95\% confidence level upper limit on the beyond-SM width of the Higgs boson
of $0.64\Gamma_H$, corresponding to
$\Gamma_H/\Gamma_H^{\rm SM} < 2.8$~\cite{CMSPAS}.
Also, a Higgs width dominated by invisible modes can be ruled out
by direct search~\cite{ATLASinvisible}.
We demonstrate here that the interference effect, because of its
different dependence on the Higgs width, allows $\Gamma_H$ to
be constrained independently of assumptions about couplings or new
decay modes.

\section{Theoretical Description}

The NLO QCD formulae for Higgs production via gluon fusion are
well known~\cite{HiggsNLO}. The SM continuum background for gluon 
fusion into two photons is also known at NLO~\cite{Bern2002jx}.
(As a component of the inclusive diphoton background, $pp\to\gamma\gamma X$,
the process $gg\to\gamma\gamma$ technically begins at
next-to-next-to-leading order (NNLO), 
but it is greatly enhanced by the large gluon parton distribution
function (PDF) at small $x$.)
Here we present the dominant NLO corrections to the interference between the 
Higgs signal and background in QCD.  

\begin{figure}[ht]
  \begin{center}
    \includegraphics[width=0.47\textwidth]{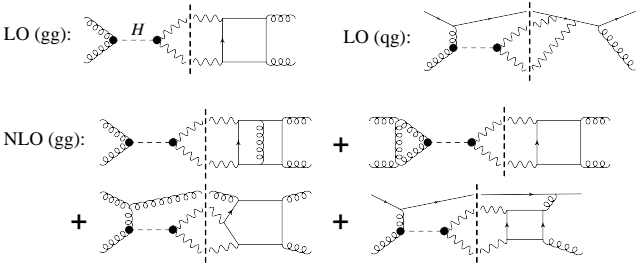}
  \end{center}
  \caption{\label{InterfFig}
   Representative diagrams for interference between the Higgs resonance
   and the continuum in the diphoton channel.  The dashed vertical lines
   separate the resonant amplitudes from the continuum ones.}
\end{figure}

Figure~\ref{InterfFig} shows, first, the leading-order (LO)
contribution to the interference [denoted by LO ($gg$)]
of the resonant amplitude $gg\to H \to \gamma\gamma$ with the one-loop
continuum $gg\to\gamma\gamma$ amplitude mediated by the five light quark
flavors.  We also include the tree-level process $qg\to \gamma\gamma q$, whose
interference with $qg\to Hq \to \gamma\gamma q$ [denoted by LO ($qg$)]
is at the same order in $\alpha_s$ as the leading $gg \to H \to \gamma\gamma$
interference, although suppressed by the smaller quark PDF.
It was already considered in refs.~\cite{deFlorian2013psa,Martin2013ula}.
The contribution from $q\bar{q}\to Hg \to \gamma\gamma g$ is numerically
tiny~\cite{deFlorian2013psa,Martin2013ula} and we will neglect it.

Finally, fig.~\ref{InterfFig} depicts the three types of
continuum amplitudes mediated by light quark loops that we include
in the dominant NLO corrections [denoted by NLO ($gg)$]:
the real radiation processes, $gg\to\gamma\gamma g$ and $qg\to\gamma\gamma q$
at one loop, and the virtual two-loop $gg\to\gamma\gamma$ process.
All these amplitudes are adapted from
refs.~\cite{Bern1993mq,Bern1994fz,Bern2001df}. 
The soft and collinear divergences in the real radiation process
are handled by dipole subtraction~\cite{Catani1996vz,Gleisberg2007md}. 
Although the contribution from $qg\to \gamma\gamma q$ via a light quark loop
is not the complete contribution to this amplitude, it forms a 
gauge-invariant subset and it is enhanced by a sum over quark flavors,
so that it gives a significant contribution to the interference at 
finite Higgs transverse momentum.

In order to parametrize possible deviations from the SM in the coupling of
the Higgs boson to the massless vector boson pairs $gg$ and $\gamma\gamma$,
we adopt the notation of ref.~\cite{Ellis2013lra} for the effective
Lagrangian,
\begin{equation}
\mathcal{L} = -\left[\frac{\alpha_s}{8\pi} c_g b_g G_{a,\mu\nu}G_{a}^{\mu\nu}
+\frac{\alpha}{8\pi} c_\gamma b_\gamma F_{\mu\nu}F^{\mu\nu}\right]\frac{h}{v} \,,
\end{equation}
where $b_{g,\gamma}$ are defined to absorb all SM contributions, 
and $c_{g,\gamma}$ differ from 1 in the case of new physics.
We divide the lineshape for the Higgs boson into a pure signal
term and an interference correction, written schematically
in the narrow-width approximation (NWA) as,
\begin{eqnarray}
\frac{d \sigma^{\rm sig}}{d M_{\gamma\gamma}} 
&=& \frac{S}{(M_{\gamma\gamma}^2-m_H^2)^2+m_H^2 \Gamma_H^2} \,, 
\label{dsigsig}\\
\frac{d \sigma^{\rm int}}{d M_{\gamma\gamma}} 
&=& \frac{(M_{\gamma\gamma}^2-m_H^2) R + m_H \Gamma_H I}
{(M_{\gamma\gamma}^2-m_H^2)^2+m_H^2 \Gamma_H^2} \,.
\label{dsigint}
\end{eqnarray} 
The signal factor $S$ is proportional to 
$c_{g}^2 c_{\gamma}^2$, while the real and imaginary parts of the interference
terms, $R$ and $I$, are proportional to $c_{g}c_{\gamma}$.  
We take the resonance mass to be $m_H = 125$~GeV and the SM width 
to be $\Gamma_H^{\rm SM} = 4$~MeV~\cite{Djouadi1997yw}. In the NWA,
the integral of the cross section over the resonance is given by 
$\pi S/(2m_H^2\Gamma_H)$ and $\pi I/(2m_H)$ for signal and
interference respectively.  Note that $R$ has a
different dependence on the Higgs width and couplings than does the integrated
signal, {\it i.e.}~$c_gc_\gamma$ versus 
$c_g^2 c_\gamma^2/\Gamma_H$.  Hence any effect due to $R$ could
be used to constrain $\Gamma_H$ independently of the Higgs couplings.

The theoretical lineshapes~(\ref{dsigsig}) and (\ref{dsigint}) are very
narrow, and strongly broadened by the experimental resolution.
The main effect of the real term $R$ after this broadening 
is to shift the apparent mass slightly~\cite{Martin2012xc}.
Following ref.~\cite{Martin2012xc}, we model the experimental
resolution by a Gaussian distribution.  
Although a definitive study of the apparent mass shift has to be performed
by the experimental collaborations, using a complete description of the
resolution and the background model, we estimate it as follows:
For the distribution in the diphoton invariant mass $\Ma$, 
the likelihood of obtaining $N$ events with 
$\Ma=\Ma_1, \Ma_2, \ldots, \Ma_N$ is,
\begin{eqnarray}
L=\frac{\L^N}{N!} e^{-\Nat} \prod^N_{i=1} \left. \sigat \right|_{\Ma=\Ma_i},
\end{eqnarray}
where $\L$ is the integrated luminosity.  Variables with tildes
denote the prediction of the ``experimental model,'' a pure Gaussian
with a variable mass parameter $\tilde{m}_H$.
For the true distribution, obtained by convoluting the sum of
eqs.~(\ref{dsigsig}) and (\ref{dsigint}) with a Gaussian
of the same width, $\sigma=1.7$~GeV, we use variables without tildes. 

To fit for the shifted mass, we minimize the test statistic $t = -2 \ln L$
with respect to $\tilde{m}_H$.  We derived the following equation
determining the mass shift $\Delta m_H \equiv \tilde{m}_H - m_H$:
\begin{eqnarray}
0 = \delta \langle t \rangle 
&\propto& \int d\Ma \frac{\sigat-\siga}{\sigat} \delta{\sigat} 
\approx \!\int\! d\Ma \frac{\sigat-\siga}{\siga} \delta{\sigat} \nonumber\\
&=& \delta \left[ \int d\Ma \frac{\left(\sigat-\siga\right)^2}{2\siga} \right]
\,,
\label{lsqfit}
\end{eqnarray}
where $\delta \equiv \delta/\delta\tilde{m}_H$. Because
$\siga$ in the denominator should include the large continuum
background, which is roughly constant throughout the range of
consideration, \eqn{lsqfit} reduces to a simple least-squares fit.
The mass shift obtained from this fit is stable once we include invariant
masses ranging out to three times the Gaussian width.

\begin{figure}[ht]
  \begin{center}
    \includegraphics[width=0.5\textwidth]{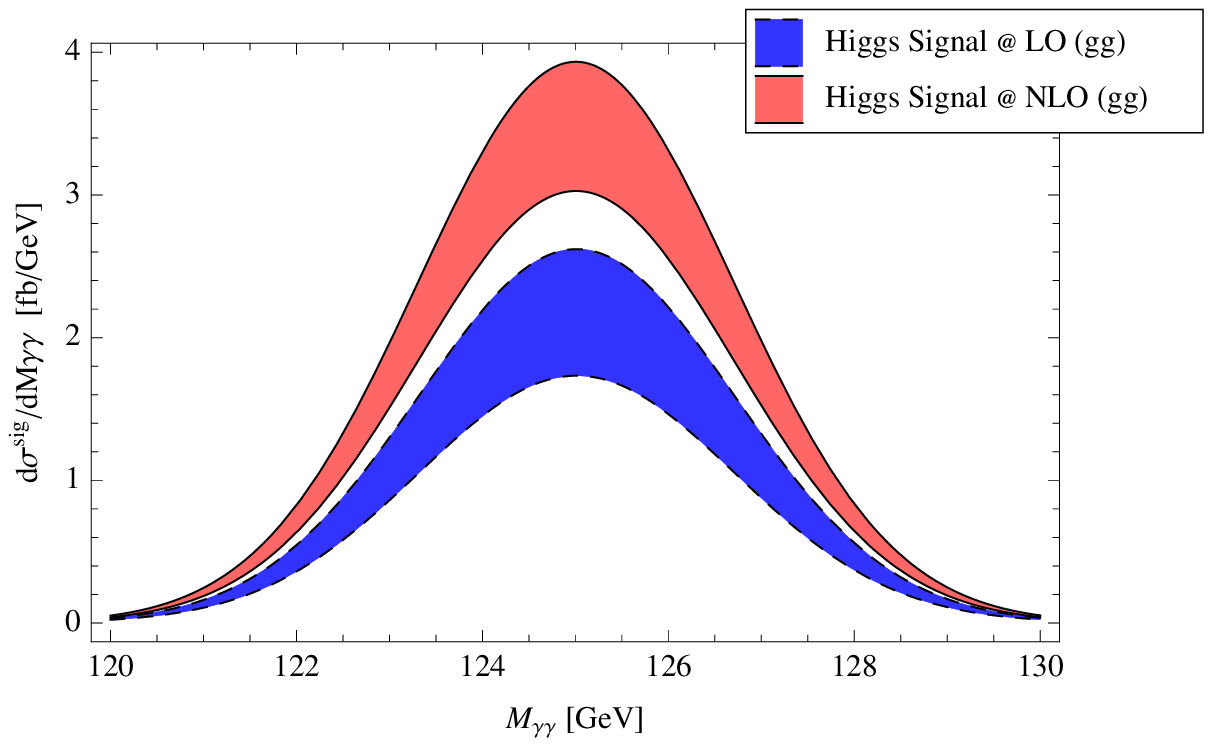}
    \includegraphics[width=0.5\textwidth]{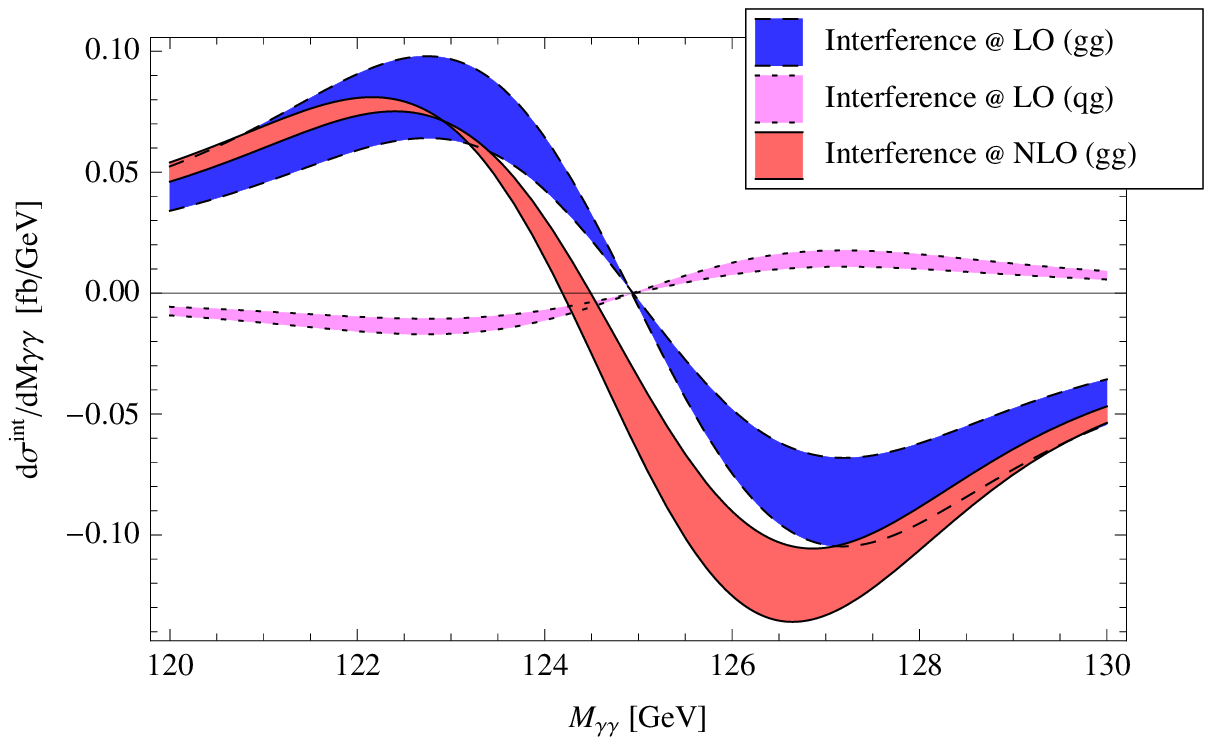}
  \end{center}
  \caption{\label{Mdistr}
   Diphoton invariant mass $M_{\gamma\gamma}$ distribution for pure signal 
   (top panel) and interference term (bottom panel) after Gaussian smearing.}
\end{figure}

The top panel of fig.~\ref{Mdistr} shows the Gaussian-smeared 
diphoton invariant mass distribution for the pure signal
at both LO and NLO in QCD.  We use the MSTW2008 NLO PDF
set and $\alpha_s$~\cite{Martin2009iq} throughout, and set
$\alpha = 1/137$.  Standard acceptance cuts are applied
to the photon transverse momenta, $p_{T,\gamma}^{\textrm{hard/soft}}>40/30$~GeV,
and rapidities, $|\eta_{\gamma}|<2.5$.  In addition, events are discarded when 
a jet with $p_{T,j}>3$~GeV is within $\Delta R_{\gamma j}<0.4$ of a photon.
A jet veto is simulated by throwing away events with $p_{T,j}>20$~GeV and 
$\eta_{j}<3$. The scale uncertainty
bands are obtained by varying $m_H/2<\mu_F,\mu_R<2m_H$
independently. Note that the NLO ($gg$) channel includes
the contribution from the $qg$ channel where the quark splits to a
gluon; this reduces dependence on the factorization scale $\mu_F$. As a
result, the scale uncertainty bands mostly come from varying the
renormalization scale $\mu_R$. 

The bottom panel of fig.~\ref{Mdistr} shows the corresponding Gaussian-smeared
interference contributions.  The contribution involving the
SM tree amplitude for $qg\to \gamma\gamma q$ is denoted by LO ($qg$).
The destructive interference from the imaginary part $I$ in \eqn{dsigint}
shows up at two-loop order in the gluon channel in the zero
mass limit of light quarks~\cite{Dixon2003yb}. It produces
the offset of the NLO ($gg$) curve from zero at $M_{\gamma\gamma} = 125$~GeV.

\begin{figure}[ht]
  \begin{center}
    \includegraphics[width=0.5\textwidth]{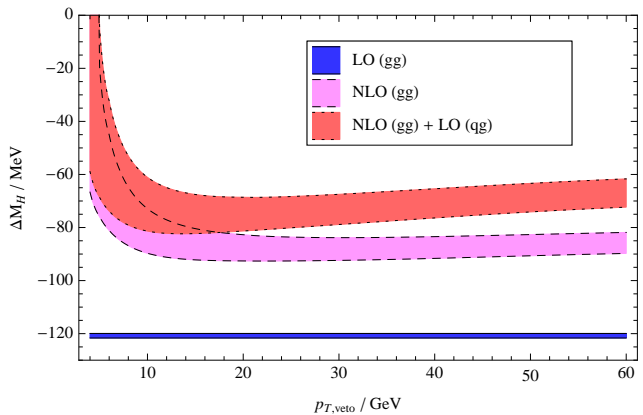}
  \end{center}
  \caption{\label{dMvspTveto}
   Apparent mass shift for the SM Higgs boson versus the jet veto $p_T$.}
\end{figure}
\begin{figure}[ht]
  \begin{center}
    \includegraphics[width=0.5\textwidth]{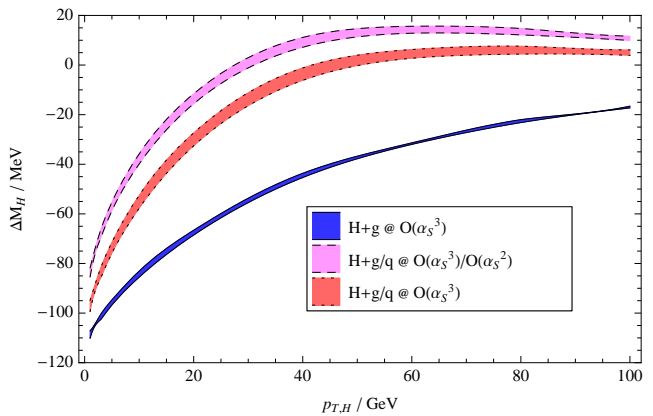}
  \end{center}
  \caption{\label{dMvspTh}
   Apparent mass shift for the SM Higgs boson versus the lower cut
   on the Higgs transverse momentum, $p_T > p_{T,H}$.}
\end{figure}

\section{Mass Shift and Width Dependence}

In fig.~\ref{dMvspTveto} we plot the apparent
Higgs boson mass shift versus the jet veto $p_T$ cut.
The mass shift for inclusive production (large $p_{T,{\rm veto}}$)
is around $70$~MeV at NLO, significantly smaller
than the LO prediction of $120$~MeV.  The reduction is mainly due to the large
NLO QCD Higgs production $K$ factor.  The $K$ factor for the SM continuum
background is also sizable due to the same gluon incoming states.
But the Higgs signal is enhanced additionally by the virtual
correction to the top quark loop, which is missing in the continuum
background~\cite{Bern2002jx}.  The $K$ factor of the interference is
between that of the signal and that of the background. This is reasonable
but not inevitable, given that only a restricted set of helicity
configurations enters the interference.
For moderate jet veto cuts, the mass shift depends very weakly on 
$p_T$ due to the smallness of the real radiation contribution.
The extra interference with quark-gluon scattering at
tree level reduces the mass shift a bit more,
as shown in the curve labeled NLO ($gg$) $+$ LO ($qg$) in 
fig.~\ref{dMvspTveto}.  At small veto $p_T$, the
results become unreliable:  large logarithms spoil the convergence of
perturbation theory, and resummation is required, which is beyond the
scope of this letter.

In fig.~\ref{dMvspTh} we remove the jet veto cut, and study how the
mass shift depends on a lower cut on the Higgs transverse momentum,
$p_T > p_{T,H}$.
This strong dependence could
potentially be observed experimentally, completely within the 
$\gamma\gamma$ channel, without having to compare against a mass
measurement using the only other high-precision channel, $ZZ^*$.
(The mass shift for $ZZ^*$ is much smaller than for $\gamma\gamma$,
as can be inferred from fig.~17 of ref.~\cite{KauerPassarino},
because $H \to ZZ^*$ is a tree-level decay, while the continuum
background $gg \to ZZ^*$ arises at one loop, the same order
as $gg\to \gamma\gamma$.)
Using only $\gamma\gamma$ events might lead to reduced experimental
systematics associated with the absolute photon energy scale.
The $p_{T,H}$ dependence of the mass shift was first studied in 
ref.~\cite{Martin2013ula}.  The dotted red band includes, in addition,
the continuum process $qg\to\gamma\gamma q$ at one loop via a light quark
loop, a part of the full ${\cal O}(\alpha_s^3)$ correction.
This new contribution partially cancels against the tree-level $qg$ channel,
leading to a larger negative Higgs mass shift.  The scale variation of the mass
shift at finite $p_{T,H}$ is very small, because it is essentially a LO
analysis; the scale variation largely cancels in the ratio between
interference and signal that enters the mass shift.

Due to large logarithms, the small $p_{T,H}$ portion of fig.~\ref{dMvspTh}
is less reliable than the large $p_{T,H}$ portion.  In using the
$p_{T,H}$ dependence of the mass shift to constrain the Higgs width,
the theoretical accuracy will benefit from using a wide first
bin in $p_T$.  One could take the difference between
apparent Higgs masses for $\gamma\gamma$ events in two bins, those 
having $p_T$ above and below, say, 40~GeV.

\begin{figure}[ht]
  \begin{center}
    \includegraphics[width=0.5\textwidth]{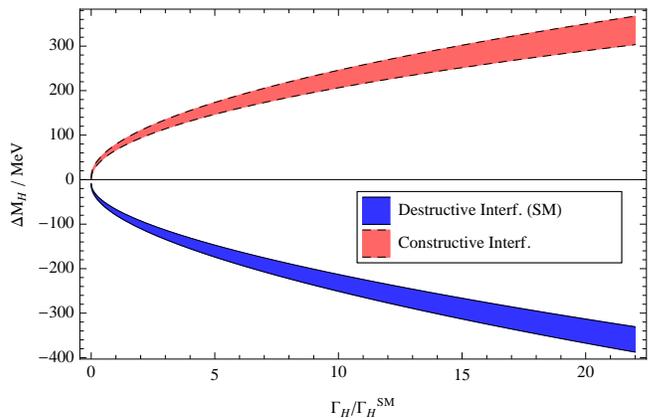}
  \end{center}
  \caption{\label{dMvsGam}
   Higgs mass shift as a function of the Higgs width.  The coupling 
   $c_{g\gamma}$ has been adjusted to maintain a constant signal strength,
   in this case $\mu_{\gamma\gamma}=1$.}
\end{figure}

Finally, we allow the Higgs width to differ from the SM prediction.
The Higgs couplings to gluons, photons, and other observed final states
should then change accordingly, in order to maintain roughly SM signal yields, 
as is in reasonable agreement with current LHC measurements. 
In particular, for the product $c_g c_\gamma = c_{g\gamma}$ entering
the dominant gluon fusion contribution to the $\gamma\gamma$ yield,
we solve the following equation,
\begin{equation}
\frac{c_{g\gamma}^2 S}{m_H\Gamma_H}+c_{g\gamma} I 
= \left( \frac{S}{m_H\Gamma_H^{\rm SM}} + I \right)
 \mu_{\gamma\gamma} \,,
\label{constyield}
\end{equation}
where $\mu_{\gamma\gamma}$ denotes the ratio of the experimental
signal strength in $gg\to H\to \gamma\gamma$ to the SM prediction
($\sigma/\sigma^{\rm SM}$).  For Higgs widths much less than 1.7~GeV,
the mass shift is directly proportional to $c_{g\gamma}/\mu_{\gamma\gamma}$.
On the right-hand side of \eqn{constyield}, the two-loop imaginary
interference term $I$ is negligible;
the fractional destructive interference in the SM is 
$m_H \Gamma_H^{\rm SM} \, I/S \approx -1.6\%$.
For $\Gamma_H \leq 100 \Gamma_H^{\rm SM} = 400$~MeV, it is a good
approximation to also neglect $I$ on the left-hand side.
Then the solution for $c_{g\gamma}$ is simply
$c_{g\gamma}=\sqrt{\mu_{\gamma\gamma} \Gamma_H/\Gamma_H^{\rm SM}}$.
Fig.~\ref{dMvsGam} plots the mass shift, assuming $\mu_{\gamma\gamma}=1$.
It is indeed proportional to $\sqrt{\Gamma_H}$ for the widths
shown in the figure, up to small corrections.
If new physics somehow reverses the sign of
the Higgs diphoton amplitude, the interference is constructive
and the mass shift is positive.  

In principle, one could apply the existing measurements
of the Higgs mass in the $ZZ^*$ and $\gamma\gamma$ channels
in order to get a first limit on the Higgs width from this method.
However, there are a few reasons why we do not do this here.
First of all, the current ATLAS~\cite{ATLASmassshift} 
and CMS~\cite{CMSmassshift} measurements are not very compatible,
\begin{eqnarray}
m_H^{\gamma\gamma} - m_H^{ZZ} &=& +2.3^{+0.6}_{-0.7}\pm 0.6~{\rm GeV~(ATLAS)} 
\nonumber\\
&=& -0.4 \pm 0.7 \pm 0.6~{\rm GeV~(CMS),}
\label{exptmassshift}
\end{eqnarray}
where the first error is statistical and the second is systematic.
Second, the experimental resolution differs from bin to bin and has
non-Gaussian tails.  Third, the precise background model can 
influence the apparent mass shift.  What we can say is that taking
$\Gamma_H = 200 \Gamma_H^{\rm SM} = 800$~MeV and neglecting the latter
factors would result in a mass shift of order 1~GeV, in the same range
as \eqn{exptmassshift}.  This is a considerably smaller width than the
first direct bound from CMS, $\Gamma_H < 6.9$~GeV at 95\% confidence
level~\cite{CMSdirectwidth}.

A measurement of $\Delta m_H$ using two $p_{T,H}$ bins in the 
$\gamma\gamma$ channel is currently limited by statistics.  At the
high luminosity LHC, with 3~ab$^{-1}$ of integrated luminosity at 14~TeV,
the statistical error on $\Delta m_H$
will drop to 50~MeV or less.  The extrapolation
of the systematic error is still somewhat uncertain,
but should result in a total error of 100~MeV or less~\cite{ATLASprivate}.
From \fig{dMvsGam}, this corresponds to a bound on the Higgs width at 
95\% C.L. that is within a factor of 15 of the SM value of 4~MeV.

\section{Summary}

In this letter, we have studied the interference of the SM Higgs boson with
the LHC diphoton continuum background at NLO in QCD. The mass shift is
largely stable for moderate jet veto $p_T$ cuts. In addition, we provide
a slightly more precise prediction for the mass shift at finite Higgs $p_T$,
by including the contribution from quark-gluon scattering via quark loops.
The strong $p_T$ dependence of the mass shift may allow its measurement
without reference to the $ZZ^*$ channel. Furthermore, we consider a scenario
in which new physics modifies the Higgs width without altering
event rates in the diphoton channel.  The mass shift increases 
rapidly with the Higgs width, which could
lead to a more direct bound on the Higgs width than
is presently available.

\section{Acknowledgment}

We are grateful to Florian Bernlochner, Glen Cowan, Daniel de Florian,
Dag Gillberg, Louis Fayard, Tom Junk, Narei Lorenzo,
Steve Martin, Jamie Saxon and Reisaburo Tanaka for helpful comments. We thank
Stefan H\"{o}che for useful discussions and help with the numerical
implementation. This research was supported by the US Department of Energy
under contract DE--AC02--76SF00515.


\end{document}